\documentclass[a4paper,aps,prd,10pt,preprintnumbers,showpacs,twocolumn,groupedaddress,superscriptaddress,nofootinbib,amsmath,amssymb]{revtex4-1}\usepackage{graphicx}\def\PRDstyle#1{#1}\def\JCAPstyle#1{}\def\Abstract#1{\begin{abstract}#1\end{abstract}}
\usepackage[utf8]{inputenc}
\usepackage[T1]{fontenc}
\usepackage{cmap}
\DeclareMathAlphabet{\pazocal}{OMS}{zplm}{m}{n}
\def\imo{i}

\def\Order#1{{\cal O}\left(#1\right)}

\begin{document}

\title{Hawking radiation of renormalization group improved regular black holes}

\JCAPstyle{
\author[1]{R.~A.~Konoplya,}\emailAdd{roman.konoplya@gmail.com}
\affiliation[1]{Research Centre for Theoretical Physics and Astrophysics, \\ Institute of Physics, Silesian University in Opava, \\ Bezručovo náměstí 13, CZ-74601 Opava, Czech Republic}
}

\PRDstyle{
\author{R. A. Konoplya}\email{roman.konoplya@gmail.com}
\affiliation{Research Centre for Theoretical Physics and Astrophysics, Institute of Physics, Silesian University in Opava, Bezručovo náměstí 13, CZ-74601 Opava, Czech Republic}
}

\Abstract{
We consider a renormalization group approach based on the idea that the primary contribution to the Schwarzschild-like black hole spacetime arises from the value of the gravitational coupling. The latter depends on the distance from the origin and approaches its classical value in the far zone. However, at some stage, this approach introduces an arbitrariness in choosing an identification parameter. There are three approaches to the identification: the modified proper length (the Bonanno-Reuter metric), the Kretschmann scalar (the Hayward metric), and an iterative, and, in a sense, coordinate-independent procedure (Dymnikova solution).
Using the WKB method, we calculated grey-body factors for the Standard Model massless test fields and their corresponding energy emission rates. For all of these solutions, we found that the intensity of Hawking radiation of massless fields is significantly suppressed by several or more orders once the quantum correction is taken into consideration. This indicates that the effect of suppression of the Hawking radiation may be appropriate to the quantum corrected black holes in asymptotically safe gravity in general and is independent on the particular choice of the identification parameter.
}

\maketitle

\section{Introduction}\label{Intro}

Quantum corrections to Hawking radiation refer to including quantum effects in the calculation of thermal radiation emitted by black holes, as originally proposed by Stephen Hawking in 1974 \cite{Hawking:1975vcx}. The original Hawking radiation was derived using semiclassical methods, treating gravity classically and considering quantum fields in a black hole background. However, when accounting for quantum effects of gravity, modifications to the Hawking radiation spectrum may occur, influencing the emission rates and energy distributions of the radiation. These corrections play a significant role in extreme conditions, particularly during the late stages of black hole evaporation, and remain an active area of research in theoretical physics and quantum gravity.

Nonetheless, it is worth investigating the evaporation of quantum-corrected black holes from a purely semi-classical approach. Notably, in the semi-classical approach used for Schwarzschild spacetime \cite{Page:1976df}, gravitons contribute less than two percent to the total massless particle emission, and even a smaller fraction when considering the emission of massive particles in the later stages of evaporation \cite{Page:1976ki}. This observation leads to the possibility of using the ordinary Hawking formula to estimate the radiation of test fields already quantized in the semi-classical approach, assuming that the contribution of gravitons to the total flux of radiation at the late stages of evaporation is relatively small. By applying this approach to the static quantum-corrected metric obtained within a specific gravitational theory, we can estimate the emission of test fields in the vicinity of the black hole.

This method has been employed in various works \cite{Konoplya:2023ahd,Bronnikov:2023uuv,Konoplya:2020jgt,Konoplya:2020cbv,Konoplya:2019ppy,Konoplya:2019hml}. In this study, we aim to pursue a similar program for the model of quantum-corrected black holes constructed using the renormalization group approach.

In this context, the construction of a model for a quantum-corrected black hole is important not only as a particular strong-gravity solution within an unknown, non-contradicting quantum gravity theory, but also because of the fundamental problem related to the final stage of Hawking evaporation and the central singularities of classical solutions.

The approach we are interested in here is based on the supposition that the gravitational coupling depends on the energy scale under consideration \cite{Bonanno:2001hi,Bonanno:2002zb,Rubano:2004jq,Pawlowski:2018swz,Koch:2013owa,Ishibashi:2021kmf,Chen:2022xjk,Platania:2019kyx} and is known as \emph{asymptotically safe gravity} (see \cite{Platania:2023srt} for a recent review). This approach is rooted in quantum field theory and the renormalization group, developing a kind of renormalization group improved analog of the Schwarzschild metric \cite{Bonanno:2000ep,Held:2019xde,Platania:2019kyx}. To work within this framework, one needs to integrate the beta function for the gravitational coupling and find the Newton constant as a function of some energy scale $k$. Then, this energy-dependent Newton constant is used in the classical black hole solution to obtain the quantum corrected lapse function, describing a regular black hole \cite{Bonanno:2000ep,Held:2019xde,Platania:2019kyx}. It is worth noting that the idea of energy dependence of the laws of physics at the level of the action is common in quantum field theory.

The gravitational coupling depends on some arbitrary renormalization group scale $k$, so the relation connecting the energy and the radial coordinate must be defined to write down the resulting quantum-corrected black hole metric. At this stage, we are aware of three opportunities for the identification of the parameter $k$:
\begin{itemize}
\item With the modified proper distance, resulting in the so-called Bonanno-Reuter black hole metric \cite{Bonanno:2000ep}.
\item With a power of the Kretschmann scalar \cite{Held:2019xde,Platania:2019kyx,Pawlowski:2018swz}, leading to the Hayward metric \cite{Hayward:2005gi}.
\item Using an iterative and coordinate-independent procedure \cite{Platania:2019kyx}, producing the Dymnikova spacetime \cite{Dymnikova:1992ux}.
\end{itemize}

Classical radiation, dominated by the so-called quasinormal modes, of the above three spacetimes has been extensively studied in a great number of papers \cite{Flachi:2012nv,Li:2013fka,Lin:2013ofa,Toshmatov:2015wga,Roy:2022rhv,Rincon:2020iwy,Panotopoulos:2020mii,Li:2013kkb,chinosII}, and accurate quasinormal modes including overtones have been recently calculated in \cite{Konoplya:2023aph,Konoplya:2022hll}. Nevertheless, to the best of our knowledge, the intensity of the quantum (Hawking) radiation of these black holes has not been considered, although general thermodynamic properties have been studied in \cite{Mandal:2022quv,Ma:2014zia,Becker:2012js}. At the same time, the Hawking temperature and, consequently, the intensity of Hawking radiation is much more sensitive to the near horizon deformations \cite{Konoplya:2023owh} than the fundamental quasinormal mode, which is exactly the case of black holes in the asymptotically safe gravity.

In the present work, we calculate grey-body factors and energy emission rates of massless test fields for all three regular black hole models in asymptotically safe gravity. It turns out that all three models are characterized by a strong suppression of Hawking radiation by a few orders of magnitude. This implies that the effect of suppression is generic in asymptotically safe gravity and does not depend on the choice of the identification parameter. Nevertheless, of the three considered models the preferable one is the coordinate independent approach by A. Platania \cite{Platania:2019kyx}, so that the qualitatively the same picture of Hawking evaporation process in all three models serves more to the justification of the other two approaches. Our results for the Hayward and Dymnikova black holes could also be used independently on the context of the asymptotically safe gravity, when restoring the initial interpretation of these metrics (balanced by outgoing radiation for the Hayward spacetime and isotropic vacuum state of matter for Dymnikova metric) and performing appropriate redefinition of constants.

Although we were limited to test fields, the emission of gravitons for these vacuum solutions of the Einstein equation could, in principle, be formally considered here as well. However, the main supposition of the renormalization group approach is that the dominant correction to the background metric is due to the running gravitational coupling, which implies the existence of other correction terms that were neglected. Under this supposition, there is no evidence that small perturbations of the black hole background spacetime will still be much smaller than the unknown terms we initially neglected in the background. Additionally, we know that for the Schwarzschild black hole, the contribution of gravitons to the total energy emission rate is quite small compared to the test fields and accounts for less than $2\%$ of the total emission.

It is worth mentioning that an alternative self-consistent deduction of the quantum-corrected black hole metric with the help of the renormalization group approach was suggested by Kazakov and Solodukhin in \cite{Kazakov:1993ha}, and the quasinormal modes and grey-body factors were analyzed for this case in \cite{Konoplya:2019xmn,Saleh:2016pke}, while the Hawking radiation was studied in \cite{Bronnikov:2023uuv}. However, the Kazakov-Solodukhin solution is not free of singularity but moves it to a finite distance from the center, and it has the same Hawking temperature as the Schwarzschild metric. This implies that the dominant quantum correction might not be extracted in that approach.
Recent study \cite{Borissova:2022mgd} suggests a Vaidya-like model of collapse producing a numerical non-static spacetime for evaporating black hole in the asymptotically safe gravity. It is interesting that estimates proposed in that model (without accurate calculations of the emission rates and consideration of the grey-body factors) also indicate that  at the later stages of evaporation the intensity of the Hawking radiation is considerably weaker than that for the Schwarzschild case (see, for example, fig. 8 in \cite{Borissova:2022mgd}).

Our paper is organized as follows: In Sec. II, we provide the basic information on the quantum-corrected black hole metrics under consideration. Sec. III is devoted to the wave-like equation for the electromagnetic and Dirac fields. Sec. IV discusses the classical scattering problem and grey-body factors. In Sec. V, we calculate the energy emission rates of the Hawking radiation. Finally, in Sec. VI, we review the results obtained in this work as well as in other papers devoted to other models of quantum-corrected black holes and mention some open problems.

\section{Renormalization group improved black hole spacetimes}\label{Classical}
Here we will briefly summarize a few basic aspects of finding quantum corrections to the Schwarzschild spacetime via the renormalization group approach.
For detailed description of the procedure we refer a reader to  \cite{Bonanno:2000ep,Held:2019xde,Platania:2019kyx}.

The obtaining of the improved Schwarzschild metric begins with the average (ghost-free) Einstein-Hilbert action in the four dimensional spacetime \cite{Bonanno:2000ep}:
\begin{equation}\nonumber
\Gamma_k[g_{\mu\nu}] = \frac{1}{16 \pi G(k)} \int \mathrm{d}^4 x \sqrt{\det(g_{\mu\nu})} \Bigl( - R(g_{\mu\nu}) + 2 \bar{\lambda}(k)\Bigl).
\end{equation}
The scale-dependent couplings are determined by the Wetterich equation
\begin{equation}
\partial_{k}\Gamma_{k} = \frac{1}{2} \text{Tr}\left[\left(\Gamma^{(2)}_{k} + \mathcal{R}_{k}\right)^{-1}\cdot\partial_{k}\mathcal{R}_{k}\right],
\end{equation}
where $\Gamma_{k}^{(2)}$ is the Hessian of $\Gamma_k$ with respect to $g_{\mu \nu}$, and $\mathcal{R}_k$ is an infrared cutoff function ~\cite{Bonanno:2000ep}.

According to \cite{Bonanno:2000ep}, the evolution of the dimensionless gravitational coupling can be described by the following relation
\begin{equation} \label{betag}
k\frac{\mathrm{d}}{\mathrm{d}k} k^2 G(k) = \left[ 2 + \frac{B_1 k^2 G(k)}{1- B_2 k^2 G(k)} \right] \: k^2 G(k),
\end{equation}
where $B_1$ and $B_2$ are some factors. The dimensionless running cosmological constant obey the inequality
$$\lambda = \bar{\lambda}/k^2 \ll 1 $$
for all scales of interest, so that one is allowed to approximate $\lambda \approx 0$ in the arguments of $B_{1}(\lambda)$ and $B_{2}(\lambda)$.

Integrating equation~\eqref{betag}, we can obtain the Newton's coupling
\begin{align} \label{basic}
G(k) &= \frac{G_0}{1 + \tilde{\omega} G_0 k^2},
\end{align}
where $\tilde{\omega}$ is a constant.
From the above relation we see that the quantum correction is essential at high energies and at the low energies the classical limit is reproduced.

The metric of a spherically-symmetric black hole has the following form
\begin{equation}\label{spherical}
\mathrm{d}s^2 = -f(r)\mathrm{d} t^2 + f(r)^{-1} \mathrm{d} r^2 + r^2 \: \mathrm{d}\Omega^2,
\end{equation}
with the only difference that now the Newton's coupling depends on $r$.

The next stage of the procedure is the identification of $k$ with some physical parameter. Three ways to identify the parameter $k$ were considered in the literature:
\begin{enumerate}
\item To identify $k(r)$ as a modified proper distance, which leads to the \emph{Bonanno-Reuter metric} \cite{Bonanno:2000ep}.
The metric function for the Bonanno-Reuter identification (supposing that in the classical limit $G_0=1$) has the form
\begin{equation}
f(r) = 1-\frac{2 M r^2}{r^3 + \frac{118}{15 \pi} \left(r + \frac{9}{2} M \right)},
\end{equation}
where $M$ is the black hole mass, measured in units of the Planck mass,
$$m_P\equiv\sqrt{\hbar c/G_0}\approx2.176\times10^{-8}kg\approx1.6\times10^{-35}m.$$
The event horizon exists when $M \geq M_{\text{crit}}$,
where
$$\frac{M_{\text{crit}}}{m_P}=\frac{23 \sqrt{5}+ 5 \sqrt{221}}{720} \sqrt{\frac{59}{3\pi}\left(31 + \sqrt{1105}\right)}\approx 3.503.$$


\item To identify $k$ as a function of the Kretschmann scalar $K$: $k(r) = \alpha K^{-1/4}$, where $\alpha = 48^{-1/4}$ \cite{Held:2019xde,Platania:2019kyx,Pawlowski:2018swz}). This identification \cite{Held:2019xde} leads to the metric coinciding, up to the redefinition of constants, with the \emph{Hayward metric} \cite{Hayward:2005gi}. The Hayward metric function has the form:
\begin{equation}
f(r) = 1-\frac{2 r^2/M^2}{r^{3}/M^{3}+ \gamma},
\end{equation}
where the event horizon exists whenever $\gamma \leq 32/27$.

\item To use a kind of coordinate independent iterative procedure for identification suggested in \cite{Platania:2019kyx}.
The first step of the iteration is the classical Schwarzschild metric, which is characterized by the parameters
\begin{equation}
k_{(0)}(r)=0 \quad\Rightarrow\quad G_{(0)}=G_0\;\;,\;\; T_{\mu\nu}^{\mathrm{eff}(0)}=0\;.
\end{equation}
The second step is defined by the replacement $G_{0} \rightarrow G(r)$, where $k=k_1(r)$ , where $k=k_1(r)$ is chosen arbitrarily and serves as an initial condition for the perturbation of the system. Consequent steps of the iteration procedure, $n>1$, are defined by the following replacement
\begin{equation}
G_{(n)}\to G_{(n+1)}(r)=\frac{G_0}{1+\tilde{\omega} G_0 \, k_{(n+1)}^2(r)}\;\;,
\end{equation}
where the cutoff function $k_{(n+1)}(r)$ is constructed as a functional of the energy-density $\rho_{(n)}(r)$ generated by the variation of $G(r)$ in the previous step
\begin{equation}\label{selfcutoff}
k_{(n+1)}^2(r)\equiv\mathcal{K}[\rho_{(n)}(r)]\;.
\end{equation}
The above procedure generates a sequence of the modified Einstein equations, admitting solutions of the form \eqref{spherical} with the lapse function $f_{(n)}(r)$.
Following~\cite{Platania:2019kyx}, we notice that in the limit $n \rightarrow \infty$ the metric function coincides with the Dymnikova black hole \cite{Dymnikova:1992ux}
\begin{equation}\label{f(r)}
f(r) = 1-\frac{2 M}{r} \left(1-e^{-\dfrac{r^3}{2 l_{cr}^2 M}}\right).
\end{equation}
Here $l_{cr}$ is a critical length-scale below which the modifications owing to the running of the Newton’s constant become negligible.
The maximal value of $l_{cr}$ allowing for existence of the event horizon is
$$l_{cr} \approx 1.138 M,$$
where $M\equiv G_0 m$ is the mass measured in units of length.
\end{enumerate}

Here we will consider Hawking radiation of test massless fields for all the three regular black hole spacetimes.

\begin{figure*}
\resizebox{\linewidth}{!}{\includegraphics{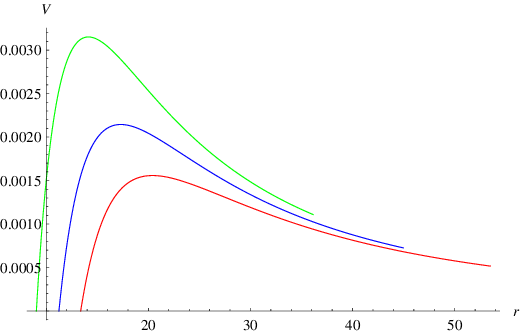}\includegraphics{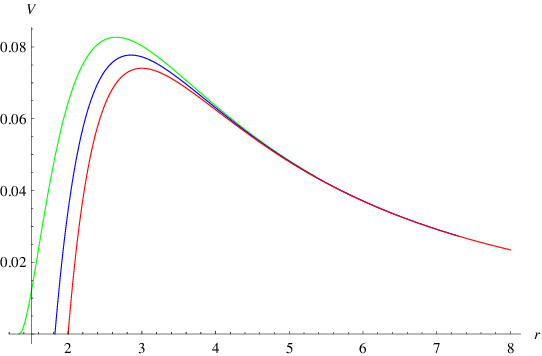}\includegraphics{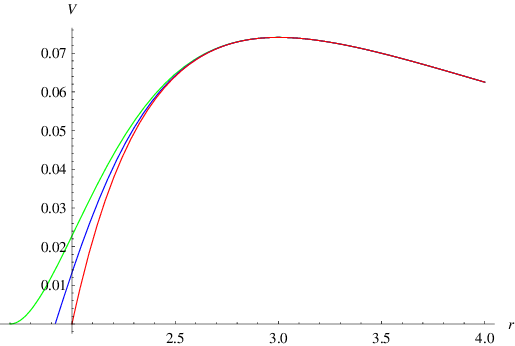}}
\caption{The effective potential for the electromagnetic field for (from left to write) a) the Bonanno-Reuter black hole: $M=5$ (top), $M=6$, $M=7$ (bottom), b) the Hayward black hole: $\gamma =32/27$ (top), $\gamma = 16/27$, $\gamma =0$ (bottom), c) the Dymnikova black hole: $l_{cr} =1.138$ (top), $l_{cr} =1.05$, $l_{cr}=0.01$ (bottom).}\label{fig0}
\end{figure*}

\section{The wave-like equations}\label{WE}
The general covariant equations for tthe electromagnetic field $A_\mu$, and the Dirac field $\Upsilon$ \cite{Brill:1957fx} are respectively written as:
\begin{subequations}\label{coveqs}
\begin{eqnarray}\label{EmagEq}
\frac{1}{\sqrt{-g}}\partial_{\mu} \left(F_{\rho\sigma}g^{\rho \nu}g^{\sigma \mu}\sqrt{-g}\right)&=&0\,,
\\\label{covdirac}
\gamma^{\alpha} \left( \frac{\partial}{\partial x^{\alpha}} - \Gamma_{\alpha} \right) \Upsilon&=&0,
\end{eqnarray}
\end{subequations}
where $F_{\mu\nu}=\partial_\mu A_\nu-\partial_\nu A_\mu$ is the electromagnetic tensor, $\gamma^{\alpha}$ are noncommutative gamma matrices and $\Gamma_{\alpha}$ are spin connections in the tetrad formalism.
After separation of the variables equations (\ref{coveqs}) take the Schrödinger-like form (see, for instance, \cite{Konoplya:2011qq,Kokkotas:1999bd} and references therein)
\begin{equation}\label{wave-equation}
\dfrac{d^2 \Psi}{dr_*^2}+(\omega^2-V(r))\Psi=0,
\end{equation}
where the ``tortoise coordinate'' $r_*$ is defined by the following relation
\begin{equation}
dr_*\equiv\frac{dr}{f(r)}.
\end{equation}

The effective potential for the electromagnetic field is
\begin{equation}\label{potentialScalar}
V(r)=f(r) \frac{\ell(\ell+1)}{r^2},
\end{equation}
where $\ell=s, s+1, s+2, \ldots$ are the multipole numbers.

For the Dirac field  we have two iso-spectral potentials
\begin{equation}
V_{\pm}(r) = W^2\pm\frac{dW}{dr_*}, \quad W\equiv \left(\ell+\frac{1}{2}\right)\frac{\sqrt{f(r)}}{r}.
\end{equation}
The iso-spectral wave functions can be transformed one into another by the Darboux transformation
\begin{equation}\label{psi}
\Psi_{+}=q \left(W+\dfrac{d}{dr_*}\right) \Psi_{-}, \quad q=const.
\end{equation}
Therefore, we will calculate quasinormal modes for only one of the effective potentials, $V_{+}(r)$, because the WKB method works better for it.

It is worth mentioning that in all three cases, the effective potentials are higher when the quantum correction is turned on,  as shown in fig. \ref{fig0} for the example of the electromagnetic field.

\section{Grey-body factors}\label{GBfactor}

\begin{figure*}
\resizebox{\linewidth}{!}{\includegraphics{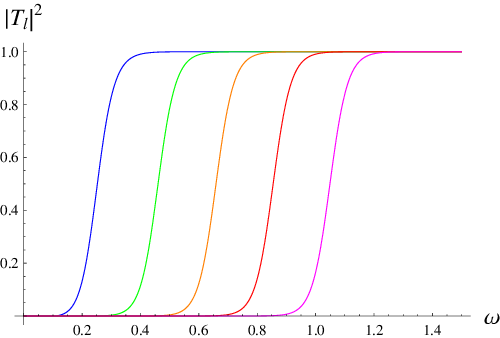}\includegraphics{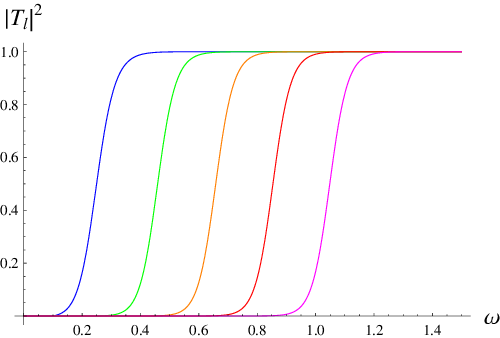}}
\caption{Grey-body factors for the Maxwell field for the Dymnikova black hole: $l_{cr} =0.5$ (left) and $l_{cr} =0.8$ (right); $M=1$. }\label{fig2a}
\end{figure*}

\begin{figure*}
\resizebox{\linewidth}{!}{\includegraphics{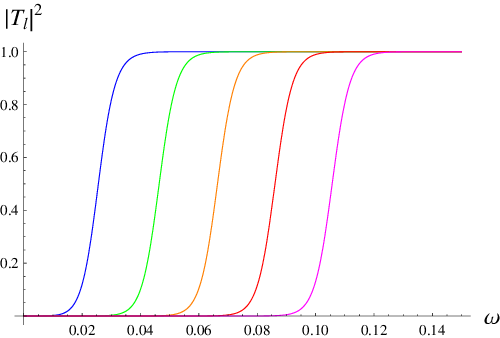}\includegraphics{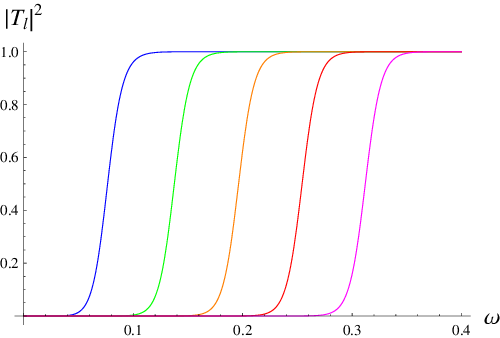}}
\caption{Grey-body factors for the Maxwell field for the Bonanno-Reuter black hole: $M =10$ (left) and $M= 3.6$ (right). }\label{fig2b}
\end{figure*}


\begin{figure*}
\resizebox{\linewidth}{!}{\includegraphics{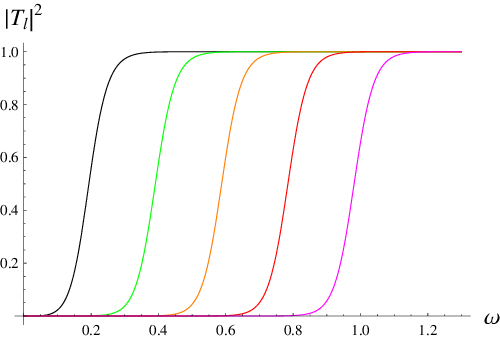}\includegraphics{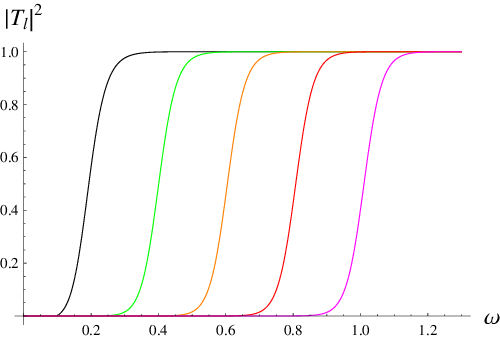}}
\caption{Grey-body factors for the Dirac field for the  Hayward black hole: $\gamma =0.5$ (left) and $\gamma= 1.1$ (right). }\label{fig2d}
\end{figure*}

Grey-body factor is a necessary ingredient for finding the proportion of the initial Hawking radiation that is reflected back to the event horizon by the potential barrier surrounding the black hole. This way, using the Hawking semi-classical formula with the grey-body factor, one can then calculate the amount of radiation that will be observed by a distant observer. As stated in \cite{Konoplya:2019ppy}, grey-body factors can be more influential than temperature.


We will examine the wave equation (\ref{wave-equation}) under the boundary conditions that allow for incoming waves from infinity. Due to the symmetry of scattering properties, this is equivalent to the scattering of a wave originating from the horizon. Thus, the boundary conditions for scattering in (\ref{wave-equation}) are as follows:
\begin{equation}\label{BC}
\begin{array}{ccll}
    \Psi &=& e^{-i\omega r_*} + R e^{i\omega r_*},& r_* \rightarrow +\infty, \\
    \Psi &=& T e^{-i\omega r_*},& r_* \rightarrow -\infty, \\
\end{array}%
\end{equation}
where $R$ and $T$ are the reflection and transmission coefficients.
\par
The effective potential for the electromagnetic field has the form of a potential barrier that decreases monotonically towards both infinities, allowing for the application of the WKB approach \cite{Schutz:1985km,Iyer:1986np,Konoplya:2003ii} to determine $R$ and $T$. As $\omega^2$ is real, the first-order WKB values for $R$ and $T$ will be real \cite{Schutz:1985km,Iyer:1986np,Konoplya:2003ii}, and
\begin{equation}\label{1}
\left|T\right|^2 + \left|R\right|^2 = 1.
\end{equation}
The potential for the Dirac field has non-monotonic behavior near the event horizon for one of the chiralities, so that, strictly speaking, there are three turning points in this case. Nevertheless, the WKB approximation is still reasonably good for this case as well.

Once the reflection coefficient is obtained, we can determine the transmission coefficient for each multipole number $\ell$
\begin{equation}
\left|{\pazocal
A}_{\ell}\right|^2=1-\left|R_{\ell}\right|^2=\left|T_{\ell}\right|^2.
\end{equation}

Here we use the higher order WKB formula \cite{Konoplya:2003ii} for a relatively accurate estimation of these transmission coefficients. However, the WKB formula is not suitable for very small values of $\omega$, which correspond to almost complete wave reflection, but, fortunately,  have negligible contributions to the overall energy emission rate. For this regime, we employed extrapolation of the WKB results at a given order to smaller $\omega$. According to \cite{Schutz:1985km,Iyer:1986np,Konoplya:2003ii}, the reflection coefficient can be expressed as follows,
\begin{equation}\label{moderate-omega-wkb}
R = (1 + e^{- 2 i \pi K})^{-\frac{1}{2}},
\end{equation}
where $K$ is determined by solving the equation
\begin{equation}
K - i \frac{(\omega^2 - V_{max})}{\sqrt{-2 V_{max}^{\prime \prime}}} - \sum_{i=2}^{i=6} \Lambda_{i}(K) =0,
\end{equation}
involving the maximum effective potential $V_{max}$, the second derivative $V_{max}^{\prime \prime}$ with respect to the tortoise coordinate, and higher order WKB corrections $\Lambda_i$. The WKB series does not guarantee convergence in each order, but only asymptotically, so that usually there is an optimal moderate order at which the accuracy is the best. This order depends on the form of the effective potential. Here we used 6th order for Maxwell perturbations and 3d order for the Dirac field with a plus-potential, because these orders provide the best accuracy in the Schwarzschild limit and the expectation is that this will also take place for the quantum corrected  black hole.

They grey-body factors per unit mass for all three cases under consideration are decreased when the quantum deformation parameter is turned on, or, in the case of the Bonanno-Reuter metric, if $M$ is decreasing. This occurs because the potential barriers are higher for quantum corrected black holes than for the Schwarzschild limit, so that higher barrier reflects bigger portion of the initial radiation.  Thus, the grey-body factors work for suppression of the Hawking radiation. In figs. (\ref{fig2a}-\ref{fig2d}) we show examples of the grey-body factors for various black holes and fields.

Using the WKB arguments \cite{Konoplya:2023moy}, we can  show analytically that the quantum corrections suppress the grey-body factors.
The first-order WKB formula gives us an expression for $K$, which we denote as $K_0$:
\begin{equation}\label{eikonalK}
-\imo K_0=\frac{\omega^2-V_0}{\sqrt{-2V_2}}.
\end{equation}
This is the eikonal formula, which can be used when the turning points are close to each other, so that the Taylor expansion near the peak of the potential barrier is well matched with the WKB expansions and the extension into the complex plane is reasonable:
\begin{equation}\label{deltaw}
\delta\equiv\omega^2-\Omega\kappa^2\approx0.
\end{equation}
Here $\Omega$ is the angular velocity of the unstable null geodesics and $\kappa = \ell + \frac{1}{2}$. Fortunately, the nontrivial (i.e. not zero and unity) values of the reflection and transmission coefficients correspond to those values of $\omega$ for which the turning points are close and $K \approx K_0 \approx 0$, which is the range of the validity of WKB approach.

Expanding $K$ near its eikonal value $K_0$ and taking $K=K_0+\Order{\kappa^{-1}}$, we find corrections to the eikonal formula (\ref{eikonalK}).
For this purpose we have
$$\frac{\omega^2}{\kappa^2}-\Omega^2\equiv\frac{\delta}{\kappa^2}\approx-\imo K_0\frac{\sqrt{-2V_2}}{\kappa^2}=\Order{\kappa^{-1}}.$$
Then, using eq. \ref{moderate-omega-wkb}, in the dominant order we find that for the Hayward black hole the squared reflection coefficient has the form
\begin{widetext}
\begin{equation}
\left|R_{\ell}\right|^2 \approx \frac{1}{\exp \left(\frac{\pi  \left(11 \gamma^2+18
   \gamma+486\right) \left(\left(10 \gamma^2+54
   \gamma+729\right) \left(\ell +\frac{1}{2}\right)^2-19683 M^2
   \omega ^2\right)}{354294 \left(\ell +\frac{1}{2}\right)}\right)+1}.
\end{equation}
For the Bonanno-Reuter black hole with an arbitrary parameter $\gamma$ instead of $118/15 \pi$ we have the following expression for the squared reflection coefficient,
\begin{equation}
\left|R_{\ell}\right|^2 \approx\frac{1}{\exp \left(-\frac{\pi  \left(1944 M^4+252 M^2 \gamma+1459
   \gamma^2\right) \left(52488 M^6 \omega ^2-486 M^4 (2 \ell
   +1)^2-270 \gamma (2 M \ell +M)^2-317 (2 \gamma \ell
   +\gamma)^2\right)}{1889568 M^8 (2 \ell +1)}\right)+1}.
\end{equation}
\end{widetext}
From the above analytical formulas it can be seen that the quantum corrections lead to the increase of the reflection coefficient and, consequently, suppression of the total emission flux to the external observer.

\section{Intensity of Hawking radiation}\label{HawkingR}

\begin{table}
\begin{center}
\begin{tabular}{|c|c|c|}
  \hline
  $M$ & Maxwell $dE/dt \cdot M^2$ & Dirac  $dE/dt \cdot M^2$  \\
\hline
  Schw ($\ell=1$) &  $0.0000335107$ &  $0.000159582$  \\
  Schw ($\ell=2$) &  $6.67916 \cdot 10^{-7}$ &  $5.86897 \cdot 10^{-6}$  \\
  Schw ($\ell=3$) &  $1.00693 \cdot 10^{-8}$ &  $1.16695 \cdot 10^{-7}$ \\
\hline
  Schw  (total) & $0.0000341888$  &  $0.000165569$ \\
\hline
  30 ($\ell=1$) &  $0.0000325103$ &  $0.000156072$ \\
  30 ($\ell=2$) &  $6.3412 \cdot 10^{-7}$ &  $5.67386 \cdot 10^{-6}$ \\
  30 ($\ell=3$) &  $9.35207 \cdot 10^{-9}$ &  $1.09462 \cdot 10^{-7}$\\
\hline
  30 (total) &   $0.0000331539$  &  $0.000161857?$  \\
\hline
  20 ($\ell=1$) &  $0.0000312768$ &  $0.000151805$ \\
  20 ($\ell=2$) &  $5.93444 \cdot 10^{-7}$ &  $5.32652 \cdot 10^{-6}$  \\
  20 ($\ell=3$) &  $8.50996 \cdot 10^{-9}$ &  $1.00874 \cdot 10^{-7}$  \\
\hline
  20 (total) &  $0.0000318789$ &  $0.000157234$  \\
  \hline
  10 ($\ell=1$) &  $0.0000249574$ &  $0.000128778$  \\
  10 ($\ell=2$) & $4.02788  \cdot 10^{-7}$ &  $2.19645  \cdot 10^{-6}$ \\
  10 ($\ell=3$) &  $4.90081 \cdot 10^{-9}$ &  $1.32767 \cdot 10^{-8}$  \\
\hline
  10 (total) & $0.0000253651$ &  $0.000132717$  \\
  \hline
  5 ($\ell=1$) & $6.32217 \cdot 10^{-6}$ & $0.0000467973$  \\
  5 ($\ell=2$) & $3.75497 \cdot 10^{-8} $ & $5.49111 \cdot 10^{-7}$  \\
  5 ($\ell=3$) & $1.65752 \cdot 10^{-10}$ &  $3.31918 \cdot 10^{-9}$  \\
\hline
  5 (total) &   $6.35989 \cdot 10^{-6}$ &  $0.0000473497$ \\
  \hline
  3.51 ($\ell=1$) &  $3.14262 \cdot 10^{-19}$ &  $1.00473 \cdot 10^{-16}$  \\
  3.51 ($\ell=2$) &  $6.11211 \cdot 10^{-23}$ &  $3.56072 \cdot 10^{-27}$  \\
  3.51 ($\ell=3$) &  $1.27276 \cdot 10^{-22}$ &  $8.61993 \cdot 10^{-37}$  \\
\hline
  3.51 (total) &  $3.1445 \cdot 10^{-19}$ &  $1.00473 \cdot 10^{-16}$  \\
 \hline
\end{tabular}
\end{center}
\caption{Bonanno-Reuter black hole: Total energy emission rate $dE/dt$ for Maxwell and Dirac particles taken with the appropriate multiplicities factors for various values of $M$. Summation over $\ell$ is done for the first five multipole moment values; The power is in units $\hbar c^{6} G^{-2} M^{-2} = 1.719 \cdot 10^{50} (M/g)^{-2} erg \cdot sec^{-1}.$}
\end{table}


\begin{table}
\begin{center}
\begin{tabular}{|c|c|c|}
  \hline
  $\gamma$ & Maxwell $dE/dt$ & Dirac  $dE/dt$  \\
\hline
  Schw ($\ell=1$) &  $0.0000335107$ &  $0.000159582$  \\
  Schw ($\ell=2$) &  $6.67916 \cdot 10^{-7}$ &  $5.86897 \cdot 10^{-6}$  \\
  Schw ($\ell=3$) &  $1.00693 \cdot 10^{-8}$ &  $1.16695 \cdot 10^{-7}$ \\
\hline
  Schw  (total) & $0.0000341888$  &  $0.000165569$ \\
\hline
  $0.05$ ($\ell=1$) &  $0.0000304085$ &  $0.000137952$ \\
  $0.05$($\ell=2$) &  $5.6591 \cdot 10^{-7}$ &  $5.25412 \cdot 10^{-6}$ \\
  $0.05$($\ell=3$) &  $7.9559 \cdot 10^{-9}$ &  $9.65131 \cdot 10^{-8}$\\
\hline
  $0.05$ (total) &   $0.0000309825$  &  $0.000143304$  \\
\hline
  $0.1$ ($\ell=1$) &  $0.0000274333$ &  $0.00012765$ \\
  $0.1$ ($\ell=2$) &  $4.74618 \cdot 10^{-7}$ &  $4.55271 \cdot 10^{-6}$  \\
  $0.1$ ($\ell=3$) &  $6.19487 \cdot 10^{-9}$ &  $7.77594 \cdot 10^{-8}$  \\
\hline
  $0.1$ (total) &  $0.0000279142$ &  $0.000132281$  \\
  \hline
  $0.25$ ($\ell=1$) &  $0.000019314$ &  $0.0000982684$  \\
  $0.25$ ($\ell=2$) & $2.60098  \cdot 10^{-7}$ &  $2.78983  \cdot 10^{-6}$ \\
  $0.25$ ($\ell=3$) &  $2.6316 \cdot 10^{-9}$ &  $3.71337 \cdot 10^{-8}$  \\
\hline
  $0.25$ (total) & $0.0000195768$ &  $0.000101096$  \\
  \hline
  $0.5$ ($\ell=1$) & $8.7512 \cdot 10^{-6}$ & $0.0000550257$  \\
  $0.5$ ($\ell=2$) & $6.64821 \cdot 10^{-8} $ & $9.19892 \cdot 10^{-7}$  \\
  $0.5$ ($\ell=3$) & $3.76036 \cdot 10^{-10}$ &  $6.93298 \cdot 10^{-9}$  \\
\hline
  $0.5$ (total) &   $8.81806 \cdot 10^{-6}$ &  $0.0000559526$ \\
  \hline
  $0.75$ ($\ell=1$) &  $2.35824 \cdot 10^{-6}$ &  $0.0000214352$  \\
  $0.75$ ($\ell=2$) &  $6.88533 \cdot 10^{-9}$ &  $1.46379 \cdot 10^{-7}$  \\
  $0.75$ ($\ell=3$) &  $1.47267 \cdot 10^{-11}$ &  $4.25797 \cdot 10^{-10}$  \\
\hline
  $0.75$ (total) &  $2.36514 \cdot 10^{-6}$ &  $0.000021582$  \\
 \hline
  $1$ ($\ell=1$) &  $1.11424 \cdot 10^{-7}$ &  $2.68168 \cdot 10^{-6}$  \\
  $1$ ($\ell=2$) &  $3.99834 \cdot 10^{-11}$ &  $2.34886 \cdot 10^{-9}$  \\
  $1$ ($\ell=3$) &  $1.04862 \cdot 10^{-14}$ &  $8.27087 \cdot 10^{-13}$  \\
\hline
  $1$ (total) &  $1.11464 \cdot 10^{-7}$ &  $2.68403 \cdot 10^{-6}$  \\
 \hline
  $31/27$ ($\ell=1$) &  $2.85062 \cdot 10^{-11}$ &  $6.6106 \cdot 10^{-8}$  \\
  $31/27$ ($\ell=2$) &  $5.71204 \cdot 10^{-16}$ &  $1.57179 \cdot 10^{-12}$  \\
  $31/27$ ($\ell=3$) &  $1.0843 \cdot 10^{-20}$ &  $4.91172 \cdot 10^{-18}$  \\
\hline
  $31/27$ (total) &  $2.85068 \cdot 10^{-11}$ &  $6.61076 \cdot 10^{-8}$  \\
 \hline
\end{tabular}
\end{center}
\caption{Hayward black hole: Total energy emission rate $dE/dt$ for Maxwell and Dirac particles taken with the appropriate multiplicities factors for various values of $\gamma$. Summation over $\ell$ is done for the first five multipole moment values; The power is in units $\hbar c^{6} G^{-2} M^{-2} = 1.719 \cdot 10^{50} (M/g)^{-2} erg \cdot sec^{-1}.$}
\end{table}

\begin{table}
\begin{center}
\begin{tabular}{|c|c|c|}
  \hline
  $l_{cr}$ & Maxwell $dE/dt \cdot M^2$ & Dirac  $dE/dt \cdot M^2$  \\
\hline
  0.5 ($\ell=1$) &  $0.000156072$ &  $0.000159577$ \\
  0.5 ($\ell=2$) &  $5.67386 \cdot 10^{-6}$ &  $5.86869 \cdot 10^{-6}$ \\
  0.5 ($\ell=3$) &  $1.09462 \cdot 10^{-7}$ &  $1.16686 \cdot 10^{-7}$\\
\hline
  0.5 (total) &   $0.0000331826$  &  $0.000165564$  \\
\hline
  0.7 ($\ell=1$) &  $0.0000311676$ &  $0.000159351$ \\
  0.7 ($\ell=2$) &  $6.1807 \cdot 10^{-7}$ &  $5.51832 \cdot 10^{-6}$  \\
  0.7 ($\ell=3$) &  $9.06041 \cdot 10^{-9}$ &  $1.06538 \cdot 10^{-7}$  \\
\hline
  0.7 (total) &  $0.0000317948$ &  $0.000164977$  \\
  \hline
  0.8  ($\ell=1$) &  $0.000026173$ &  $0.000139663$  \\
  0.8 ($\ell=2$) & $4.50861 \cdot 10^{-7}$ &  $4.44441 \cdot 10^{-6}$ \\
  0.8 ($\ell=3$) &  $5.74457 \cdot 10^{-9}$ &  $7.15587 \cdot 10^{-8}$  \\
\hline
  0.8 (total) & $0.0000266296$ &  $0.00014418$  \\
  \hline
  0.9 ($\ell=1$) & $0.0000157672$ & $0.0000991748$  \\
  0.9 ($\ell=2$) & $1.83205 \cdot 10^{-7} $ & $2.21015 \cdot 10^{-6}$  \\
  0.9 ($\ell=3$) & $1.56506 \cdot 10^{-9}$ &  $2.33984\cdot 10^{-8}$  \\
\hline
  0.9 (total) &   $0.0000159519$ &  $0.000101409$ \\
  \hline
  1 ($\ell=1$) &  $0.0000495318$ &  $0.0000218637$  \\
  1 ($\ell=2$) &  $1.72361 \cdot 10^{-8}$ &  $3.07147 \cdot 10^{-7}$  \\
  1 ($\ell=3$) &  $1.09833 \cdot 10^{-10}$ &  $1.44918 \cdot 10^{-9}$  \\
\hline
  1 (total) &  $0.0000495492$ &  $0.0000221723$  \\
    \hline
  1.1 ($\ell=1$) &  $5.70229 \cdot 10^{-8}$ &  $1.36432 \cdot 10^{-6}$  \\
  1.1 ($\ell=2$) &  $5.45696 \cdot 10^{-12}$ &  $1.71802 \cdot 10^{-8}$  \\
  1.1 ($\ell=3$) &  $5.48403 \cdot 10^{-16}$ &  $1.08041 \cdot 10^{-10}$  \\
\hline
  1.1 (total) &  $5.70283 \cdot 10^{-8}$ &  $1.38161 \cdot 10^{-6}$  \\
 \hline
\end{tabular}
\end{center}
\caption{Dymnikova black hole: Total energy emission rate $dE/dt$ for Maxwell and Dirac particles taken with the appropriate multiplicities factors for various values of $l_{cr}$. Summation over $\ell$ is done for the first five multipole moment values; The power is in units $\hbar c^{6} G^{-2} M^{-2} = 1.719 \cdot 10^{50} (M/g)^{-2} erg \cdot sec^{-1}.$}
\end{table}

\begin{figure*}
\resizebox{\linewidth}{!}{\includegraphics{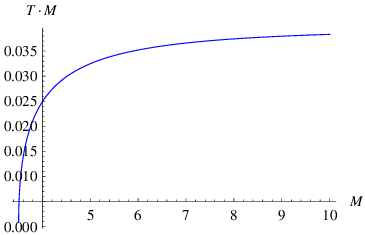}\includegraphics{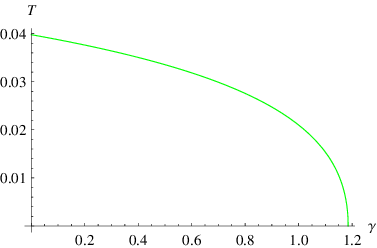}\includegraphics{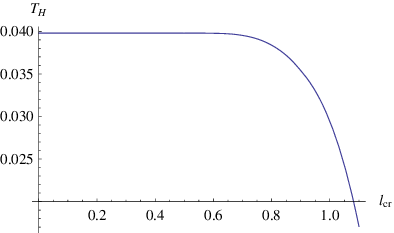}}
\caption{Temperature for the Bonanno-Reuter (left), Hayward and Dymnikova (right) black holes as a function of the corresponding quantum deformation.}\label{fig1}
\end{figure*}

In the following analysis, we will make the assumption that the black hole is in a state of thermal equilibrium with its surroundings. This means that the temperature of the black hole remains constant between the emission of two consecutive particles. According to this assumption, the system can be characterized by the canonical ensemble, which is extensively discussed in the literature (for example, see \cite{Kanti:2004nr}). Consequently, the well-known formula for the energy emission rate of Hawking radiation \cite{Hawking:1975vcx} can be applied:
\begin{align}\label{energy-emission-rate}
\frac{\text{d}E}{\text{d} t} = \sum_{\ell}^{} N_{\ell} \left| \pazocal{A}_l \right|^2 \frac{\omega}{\exp\left(\omega/T_\text{H}\right)\pm1} \frac{\text{d} \omega}{2 \pi},
\end{align}
were $T_H$ is the Hawking temperature, $A_l$ are the grey-body factors, and $N_l$ are the multiplicities, which depend on the number of species of particles and $\ell$.
The Hawking temperature is \cite{Hawking:1975vcx}
$$
T = \left.\frac{f'(r)}{(4 \pi)}\right\vert_{r=r_{H}},
$$
where $f(r)$ is the laps metric function and $r_{H}$ is radius of the event horizon. The dependence of the temperature on the quantum deformation parameter is shown in fig. \ref{fig1}. There one can see that the Hawking temperature is decreasing when the quantum deformation grows for all the three models.

If a black hole is small enough, or, equivalently, the quantum deformation is sufficiently large, then massive particles, such as electrons and positrons, are emitted  ultra-relativistically, that is, roughly at the same rate for each helicity as massless neutrinos. Here we will consider only massless particles of the Standard Model and neglect the emission of gravitons. As mentioned before,  the reason for the latter is based on fact that gravitons contribute usuallyrelatively small fraction of the total emission flow in the Schwarzschild limit and one could expect that the matter fields will contribute the most part of radiation also for the quantum correct black hole.

\begin{figure*}
\resizebox{\linewidth}{!}{\includegraphics{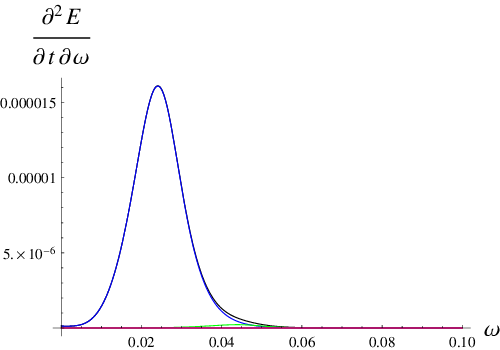}\includegraphics{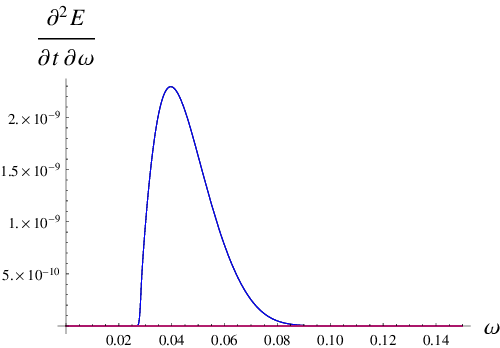}}
\caption{Energy emission rate per unit of frequency for the Maxwell field for the Bonanno-Reuter black hole: $M =10$ (left) and $M= 3.6$ (right).}\label{fig3a}
\end{figure*}


\begin{figure*}
\resizebox{\linewidth}{!}{\includegraphics{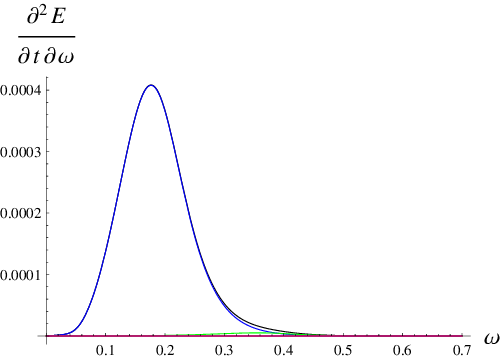}\includegraphics{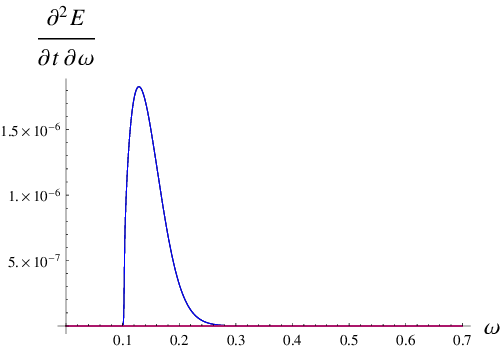}}
\caption{Energy emission rate per unit of frequency for the Dirac field for the Hayward black hole: $\gamma =0.5$ (left) and $\gamma= 1.1$ (right).}\label{fig3c}
\end{figure*}

\begin{figure*}
\resizebox{\linewidth}{!}{\includegraphics{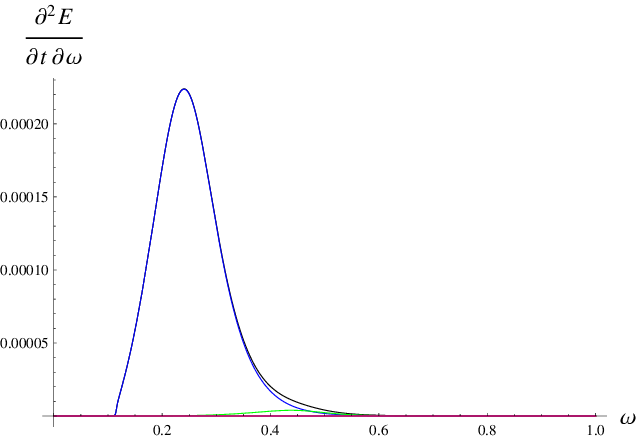}\includegraphics{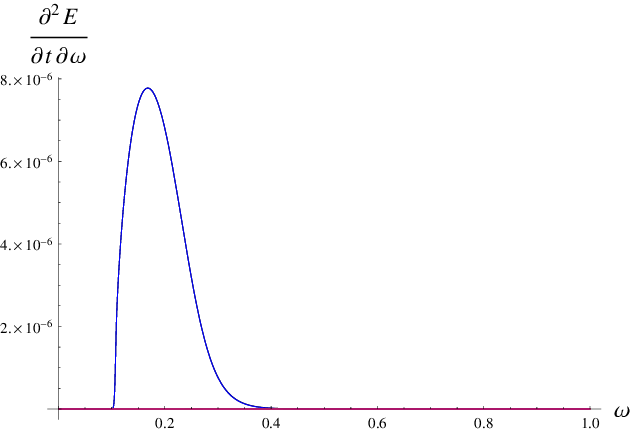}}
\caption{Energy emission rate per unit of frequency for the Maxwell field for the Dymnikova black hole: $l_{cr} =0.5$ (left) and $l_{cr} =0.8$ (right); $M=1$. }\label{fig3d}
\end{figure*}

The multiplicity factors for the four dimensional spherically symmetrical black holes case consists from
the number of degenerated $m$-modes (which are $m = -\ell, -\ell+1, ....-1, 0, 1, ...\ell$, that is  $2 \ell +1$ modes) multiplied by the number of species of particles which depends also on the number of polarizations and helicities of particles. Therefore, for the Maxwell and Dirac fields the  multiplicity factors $N_{\ell}$  have the form
\begin{equation}
N_{\ell} = 2 (2 \ell+1) \qquad (Maxwell),
\end{equation}
\begin{equation}
N_{\ell} = 8 \ell \qquad (Dirac),
\end{equation}
The multiplicity factor for the Dirac field is fixed taking into account both the ``plus'' and ``minus'' potentials which are related by the Darboux transformations, what leads to the iso-spectral problem and the same grey-body factors for both chiralities. We will use here the ``minus'' potential, because the WKB results are more accurate for that case in the Schwarzschild limit.

The energy emission rates per unit of frequency are shown in Figs. (\ref{fig3a}-\ref{fig3d}). From the figures, it can be observed that as the quantum correction grows, the emission at smaller frequencies is suppressed for all three models. However, for the same values of the black hole mass $M$, the peak of the energy emission rate occurs at approximately the same frequency $\omega$.

The total energy emission rate, after summation over all frequencies, is presented in Tables I-III for the Bonanno-Reuter, Hayward, and Dymnikova black holes, respectively. From the tables, it is evident that the energy emission rate for the Bonanno-Reuter black hole drastically decreases by 15 orders for bosons and 11 orders for fermions. Similarly, for black holes with moderate quantum deformation, around $M \sim 30$, there is a noticeable decrease in emission.
In the case of the Hayward black hole, characterized by the deformation parameter $\gamma$, a decrease of 7 orders for bosons and 4 orders for fermions is observed (see Table II). Notably, when the deformation parameter $\gamma$ is far from its extreme limit corresponding to the Planck scales, there is a considerable suppression of Hawking radiation, reaching tens of percents. Additionally, this model exhibits the highest asymmetry between bosons and fermions. Finally, for the Dymnikova black holes, a suppression of 4 orders for bosons and 2 orders for fermions is observed. However, we encountered challenges in accurately studying the extreme limit in these models. In summary, all three models are characterized by a formidable suppression of the energy emission rate, with bosons being more suppressed than fermions across all of them. Moreover, this suppression is noticeable even at moderate quantum deformations, far from the Planck scale

This effect occurs due to two factors. The first and dominant factor is the reduction in the Hawking temperature when considering quantum corrections. The second, though less important, factor is the diminished grey-body factors due to higher potential barriers around black holes.

Note that we have not considered the emission of massive particles here, which are mostly emitted at the later stages of evaporation. However, simple arguments suggest that similar suppression must occur for massive particles as well. The potential barriers for massive particles are again higher for the quantum-corrected black hole compared to the Schwarzschild case, leading to more significant reflection and consequently smaller grey-body factors. At the same time, the temperature will work to suppress radiation in the same way.

The energy emitted causes the black-hole mass to decrease at the following rate \cite{Page:1976df}
\begin{equation}
\frac{d M}{d t} = -\frac{\hbar c^4}{G^2} \frac{\alpha_{0}}{M^2},
\end{equation}
where we have restored the dimensional constants. Here $\alpha_{0} = d E/d t$ is taken for a given initial mass $M_{0}$. As the black hole is assumed to be only slightly heavier than the Planck mass, we cannot assume that it spends most of its time near an original state $M_0$.  Integrating the equation above would not provide a correct estimate of the black hole's lifetime. Thus, within our approach, when considering sufficiently strongly deformed black holes, we rely upon the particular energy emission rates, rather than averaging over periods of time. This is because as the black hole evaporates and approaches the Planck scale, the corresponding quantum deformation parameter is implied to change towards its extreme value.

\section{Conclusions}

Here we have calculated grey-body factors and energy emission rates for massless particles emitted in the vicinity of quantum-corrected black holes in asymptotically safe gravity. Since the concrete model of the quantum-corrected black hole is not unique but depends on the identification of the cut-off parameter, we have considered all known black hole models within this approach, namely the Bonanno-Reuter, Hayward, and Dymnikova metrics. Interestingly, the picture of Hawking evaporation is similar for all three models, suggesting that the observed features are independent of the particular choice of identification. Notably, we found that the energy emission rate is greatly suppressed overall, especially at small frequencies.

An important question in this context is whether the effect of the suppression of Hawking radiation is a general feature or depends on the renormalization group approach to quantum corrections. Is it only applicable to asymptotically safe gravity or independent of the method used to construct quantum corrections to the black hole spacetime? To answer this, we can compare it with other approaches to quantum corrections to Einstein gravity. Fortunately, Hawking radiation has been recently analyzed for many of these approaches.

In the majority of cases, modifications to Einstein gravity in four or higher dimensions result in a suppression effect. For example, in Einsteinian cubic gravity, the Maxwell field is suppressed by up to two orders, while the Dirac field experiences a suppression of approximately one order \cite{Konoplya:2020jgt}. In the $4D$-Einstein-Gauss-Bonnet theory, suppression occurs by a few times even at small values of the coupling constant \cite{Konoplya:2020cbv}. Einstein-Weyl gravity shows a slight enhancement at a small coupling, attributed to larger grey-body factors, but it experiences considerable suppression at higher coupling, where the temperature plays a primary role \cite{Konoplya:2019ppy}. The evaporation of the Bardeen black hole is characterized by a few orders of suppression, as demonstrated in \cite{Konoplya:2023ahd}. Finally, it is worth noting that in higher dimensional Einstein-Gauss-Bonnet theory, there is a few orders of suppression when compared with the $D$-dimensional Einstein theory \cite{Konoplya:2010vz}. In the case of apparent enhancement in the Einstein-dilaton-Gauss-Bonnet theory \cite{Konoplya:2019hml}, a comparison with the Schwarzschild case must be made not in units of the radius of the event horizon, but appropriately rescaled to the units of mass.

It seems that the only evidently opposite effect has been observed in the non-perturbative Kazakov-Solodukhin model of quantum correction \cite{Kazakov:1993ha} \cite{Bronnikov:2023uuv}. However, in this model, the Hawking temperature of the quantum-corrected black hole remains the same as in the Schwarzschild case, which raises some skepticism regarding the possibility that the calculated correction is dominant. The above observations indicate that quantum corrections may indeed lead to the suppression of Hawking radiation. Nevertheless, it is essential to acknowledge that a final, single, and non-contradictory model for quantum black holes has not yet been established. Further research and theoretical developments are still needed in this area.

\begin{acknowledgments}
R. K. would like to acknowledge useful discussions with Alexander Zhidenko and Alfio Bonanno and the Istituto Nazionale di Astrofisica of Catania University for hospitality.
\end{acknowledgments}

\bibliographystyle{unsrt}
\bibliography{Bibliography}

\end{document}